\documentclass[letterpaper]{jpconf}
\usepackage{graphicx}
\newcommand{\ba}{\begin{eqnarray}}
\newcommand{\ea}{\end{eqnarray}}

\begin{document}

\title{Strangeness suppression in the unquenched quark model}

\author{Roelof Bijker}
\address{Instituto de Ciencias Nucleares, 
Universidad Nacional Aut\'onoma de M\'exico, 
A.P. 70-543, 04510 M\'exico, D.F., M\'exico}
\ead{bijker@nucleares.unam.mx}

\author{Hugo Garc{\'{\i}}a-Tecocoatzi}
\address{Instituto de Ciencias Nucleares, 
Universidad Nacional Aut\'onoma de M\'exico, 
A.P. 70-543, 04510 M\'exico, D.F., M\'exico}
\address{I.N.F.N., Sezione di Genova, via Dodecaneso 33, I-16146 Italy}
\ead{hugo.garcia@nucleares.unam.mx}

\author{Elena Santopinto}
\address{I.N.F.N., Sezione di Genova, via Dodecaneso 33, I-16146 Italy}
\ead{elena.santopinto@ge.infn.it}

\begin{abstract}
In this contribution, we discuss the strangeness suppression in the proton in the framework 
of the unquenched quark model. The theoretical results are in good agreement with the values 
extracted from CERN and JLab experiments. 
\end{abstract}

\section{Introduction}

The importance of multiquark configurations in the proton such as $qqq-q\bar{q}$ has been inferred 
from the observed violation of the Gottfried sum rule, that is that there are more anti-d than anti-u  
quarks in the proton, and the proton spin crisis in which only about one third of the proton spin is carried 
by the quark spins. Recently, interesting new results were reported by the CLAS Collaboration \cite{Mestayer}, 
in which the strangeness suppression in the proton was determined from the production rates of baryon-meson 
states in exclusive reactions, {\it i.e.} without the production of an intermediate baryon resonance.  

The purpose of this contribution is study the observed production rates of meson-baryon final states 
and the strangeness suppression in the proton in the framework of an unquenched quark model. 

\section{Strangeness suppression} 

In the unquenched quark model (UQM) the effects of quark-antiquark pairs are introduced via a $^{3}P_{0}$ 
mechanism \cite{Tornqvist,Zenczykowski,Geiger,uqm1,uqm2,uqm3}. The production ratios of baryon-meson states 
can be expressed as the product of a spin-flavor-isospin factor and a radial integral \cite{plb}
\begin{equation}
\frac{p \rightarrow \Lambda K^+}{p \rightarrow n \pi^+} = 
\frac{27}{50} \frac{I_{N \rightarrow \Lambda K}}{I_{N \rightarrow N \pi}} ~,
\end{equation}
with
\begin{equation}
I_{A \rightarrow BC} = \int_{0}^{\infty} \frac{k^4 \mbox{e}^{-2F^2k^2}}{\Delta E_{A \rightarrow BC}^2(k)} dk ~. 
\end{equation}
The energy denominator represents the energy difference between initial and final hadrons calculated in the 
rest frame of the initial baryon $A$. The value of $F^2$ depends on the size of the harmonic oscillator wave 
functions for baryons and mesons and the Gaussian smearing of the pair-creation vertex, and is taken from  
\cite{uqm3} to be $F^2=2.275$ GeV$^{-2}$. Table~\ref{tab:res} shows that the observed rates are 
reproduced very well by our calculation. The calculated ratio $p \rightarrow p \pi^0 / p \rightarrow n \pi^+ = 1/2$ 
is a consequence of the isospin symmetry of the UQM. 

\begin{table}[t]
\begin{center}
\caption{Ratios of electroproduction cross sections.}
\label{tab:res}  
\vspace{15pt}
\begin{tabular}{ccc} 
\hline 
\hline 
\noalign{\smallskip}
Ratio                    &  UQM \cite{plb} & Exp. \cite{Mestayer} \\ 
\noalign{\smallskip}
\hline 
\noalign{\smallskip}
$p \rightarrow \Lambda K^+/p \rightarrow n\pi^+$     & 0.227 & $0.19 \pm 0.01 \pm 0.03$ \\  
$p \rightarrow \Lambda K^+/p \rightarrow p\pi^0$     & 0.454 & $0.50 \pm 0.02 \pm 0.12$ \\  
$p \rightarrow p\pi^0/p \rightarrow n\pi^+$          & 0.500 & $0.43 \pm 0.01 \pm 0.09$ \\
\noalign{\smallskip}
\hline 
\hline
\end{tabular}
\end{center}
\end{table}

In Ref.~\cite{Mestayer} a simple factorization model was proposed to extract the ratio of $q\bar{q}$ 
probabilities. In the UQM the corresponding ratios for exclusive two-body production can be determined in 
a straightforward way. The present calculation takes into account all channels involving pseudoscalar mesons 
($\pi$, $K$, $\eta$ and $\eta'$) in combination with octet and decuplet baryons.  
The results are presented in Table~\ref{strange}. The value for the strangeness suppression 
$\lambda_s=2s\bar{s}/(u\bar{u}+d\bar{d})$ is in good agreement with both the values determined 
in exclusive reactions \cite{Mestayer} and in high-energy production \cite{Bocquet}. 

\begin{table}[t]
\begin{center}
\caption{Strangeness suppression in the proton.}
\label{strange}
\vspace{15pt}
\begin{tabular}{cccc}
\hline
\hline
\noalign{\smallskip}
Ratio                           & UQM \cite{plb} & Exp. & Ref. \\
\noalign{\smallskip}
\hline
\noalign{\smallskip}
$s\bar{s}/d\bar{d}$             & $0.265$ & $0.22 \pm 0.07$ & \cite{Mestayer} \\
$u\bar{u}/d\bar{d}$             & $0.568$ & $0.74 \pm 0.18$ & \cite{Mestayer} \\ 
$2s\bar{s}/(u\bar{u}+d\bar{d})$ & $0.338$ & $0.25 \pm 0.08$ & \cite{Mestayer} \\ 
                                &         & $0.29 \pm 0.02$ & \cite{Bocquet} \\
\noalign{\smallskip}
\hline
\hline
\end{tabular}
\end{center} 
\end{table}

\section{Summary and conclusions}

In conclusion, the observed ratios for the production of baryon-meson channels in exclusive reactions 
can be understood in a simple and transparent way in the framework of the UQM. It is important to 
emphasize that the UQM results do not depend on the strength of the $^{3}P_0$ quark-antiquark 
pair creation vertex, The value of the remaining coefficient ($F^2$) was taken from previous work, 
no attempt was made to optimize their values. Finally, the UQM value for the strangeness suppression 
in the proton is in good agreement with the value determined in exclusive reactions \cite{Mestayer} 
as well as the result from high-energy production \cite{Bocquet}. 

\ack

This work was supported in part by research grant IN107314 from PAPIIT-DGAPA, Mexico 
and in part by INFN, Sezione di Genova, Italy.

\end{document}